\documentclass[aps,prb,twocolumn,showpacs,showkeys,floadfix,groupedaddress]{revtex4-1}
\usepackage{bm}
\usepackage{bbm}
\usepackage{amsmath}
\usepackage{amssymb}
\usepackage{t1enc}
\usepackage[cp1250]{inputenc}
\usepackage[english]{babel}
\usepackage{graphicx}

\begin{document}

\title{Electronic and hole spectra of layered systems\\ of cylindrical rod arrays: solar cell application}

\author{J. W. Kłos}
\affiliation{Surface Physics Division, Faculty of Physics, Adam Mickiewicz
University,\\
Umultowska 85, 61-614 Poznań, Poland}
\email{klos@amu.edu.pl}
\author{M. Krawczyk}
\affiliation{Surface Physics Division, Faculty of Physics, Adam Mickiewicz
University,\\
Umultowska 85, 61-614 Poznań, Poland}

\begin{abstract}
We have computed the electronic and hole spectra of a 3D superlattice consisting of layers of GaAs rods of finite height arranged in a hexagonal lattice and embedded in an ${\rm Al_{d}Ga_{1-d}As}$ matrix, alternating with spacer layers of homogeneous AlAs. The spectra are calculated in the envelope function approximation, with both light-hole and heavy-hole subbands and hole spin degeneracy taken into account. The application of thick spacers allows to investigate the band structure of isolated layers of cylindrical rods. We estimate the ultimate efficiency of solar energy conversion in a solar cell based on an array of cylindrical quantum dots versus the dot height, and determine the optimal value of this parameter.
\end{abstract}

\pacs{%
73.21.Cd,
84.60.Jt
}

\keywords{%
intermediate band solar cell,
superlattices,
energy spectrum}

\maketitle

\section{Introduction}
Systems with intermediate energy bands between the valence band and the conduction band offer an increased efficiency of conversion of solar energy into electricity\cite{3g2}. The theoretical ultimate efficiency limit in solar cells with a single intermediate\cite{luque,green} band is about 68\%\cite{klos2}, much above the ultimate efficiency limit, estimated at 42\%\cite{SQ}, in systems without intermediate band.

One of the methods used for generating extra energy bands consists in introducing an additional periodicity exceeding that of the crystal lattice. This is used in semiconductor superlattices\cite{tsaka, shao, green2, tomic, fan,klos2,krawczyk}, in which the system is a heterostructure with dots, rods or layers of a semiconductor material arranged periodically in a matrix of another material. If the period of the superlattice exceeds substantially that of the crystal lattice, the envelope function approximation is applicable. In this case, the parabolic conduction band bottom and valence band top are assumed to be split into a number of energy minibands.

In ${\rm Al_{d}Ga_{1-d}As}$, an alloy of zincblende structure, the valence band top is composed of bands of light and heavy holes, in which the spin degeneracy is lifted for certain nonzero wave vectors ($k_{z}\neq 0$). Thus, the structure of minibands within the valence band is rather complex\cite{datta}. Since the values of effective mass of light and heavy holes are relatively high compared to those of electrons, minibands and minigaps in the valence band are much narrower than the gap between the conduction band and the valence band. The structure of minibands in the conduction band is more transparent because of the occurrence of only one electronic band (in a monolithic material) split into relatively large minigaps and minibands in the superlattice. 

For typical values of lattice constant of the superlattice the lowest conduction miniband, lying below the matrix material potential (i.e. in the potential well generated by the dots or rods), is well detached from the higher minibands, which tend to overlap. In our estimation of solar cell efficiency the lowest conduction miniband is assumed to act as an intermediate band (IB), while the minibands in the conduction band (CB) above it, as well as the valence band (VB) as a whole, form continuous blocks. The ultimate efficiency determined in this study refers to this type of band structure. The ultimate efficiency reflects how the band structure of the solar cell matches the solar spectrum, with inevitable losses in the conversion of light quanta into electricity due to the incomplete absorption of photons (quanta of energy below the width of the narrowest bandgap are not absorbed in the system) and thermalization processes (a part of energy of photons above the bandgap width is dissipated in thermal contact with the lattice).

The objective of this study is to determine the electronic and hole spectra for an isolated array of cylindrical quantum dots arranged in a hexagonal lattice. We investigate the superlattice spectrum versus the quantum dot layer thickness (the height of the cylindrical dots). Next, we calculate the ultimate efficiency of a solar cell based on the considered structure and examine its dependence on the quantum dot layer thickness.

The paper is organized as follows. After this Introduction we discuss the geometry of a 3D superlattice of cylindrical dot arrays. Next, we present the effective mass approximation applied to the description of electronic and hole states in a heterostructure based on AlGaAs. Our results, showing spectra of quantum dot arrays and the ultimate efficiency versus the array thickness, are discussed in a separate section. This is followed by Conclusions, which close the paper.

\section{Superlattice geometry}
\begin{figure}[!tbp]
\centering
\includegraphics[keepaspectratio, width = 3in]{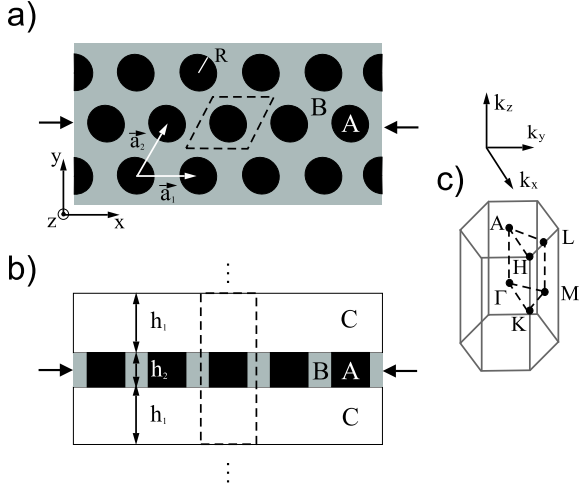}
\caption{\label{fig:geometry} Structure of three-dimensional semiconductor superlattice in cross section (a) perpendicular and (b) parallel to rod axes. Thick arrows in (a) and (b) indicate the position of cross section planes. Arrays of rods (bleack area, material A) embedded in matrix (gray area, material B) are separated by spacer layers (white area, material C). Dashed lines indicate unit cell limits (in real space); the first BZ and the irreducible part of it, is presented in (c). $R$ is a cylindrical rod radius, $\vec{a}_{1}$ and $\vec{a}_{2}$ are hexagonal lattice vectors ($|\vec{a}_{1}|=|\vec{a}_{1}|=a$  ).}
\end{figure}
Let us consider a 3D lattice composed of layers of cylindrical quantum dots arranged in a hexagonal lattice, as shown in Fig.\ref{fig:geometry}. Acting as quantum wells, the quantum dots (made of GaAs) are embedded in a matrix material (${\rm Al_{0.35}Ga_{0.65}As}$), which represents a potential barrier for both electrons and holes. The layers of quantum dots (quantum dot arrays) alternate with homogenous spacer layers of AlAs. Thick spacer layers having a high potential barrier of AlAs imply the  isolation of individual dot layers.

\section{Model}
We use the effective mass approximation to calculate the electronic and hole states, and assume the electronic bands do not interact with the bands of light and heavy holes. The calculation of the electronic spectrum is based on the Ben Daniel Duke hamiltonian with space-variable effective mass $m^{*}$ and variable position of the conduction band bottom $E_{C}$\cite{duke}:
\begin{eqnarray}
      \left[-\alpha\left(\frac{\partial}{\partial x}\frac{1}{m^{*}({\bm r})}\frac{\partial}{\partial x}+
\frac{\partial}{\partial y}\frac{1}{m^{*}({\bm r})}\frac{\partial}{\partial y}+
\frac{\partial}{\partial z}\frac{1}{m^{*}({\bm r})}\frac{\partial}{\partial z}
\right)\right.\nonumber\\\Big.+E_{C}({\bm r})\Big]\Psi_{e}({\bm r})=E\Psi_{e}({\bm r}),   \label{eq:duke}
\end{eqnarray}
where $\alpha=\hbar/2m_{e}$, $m_{e}$ denotes the free electron mass and $\Psi_{e}$ is an envelope of the electron function. The determination of hole states requires taking into account both the light-hole and heavy-hole bands. Since, in general, the hole spin degeneracy is lifted in a 3D system, four components of the hole envelope function ${\bm \Psi}_{h}$ must be taken into consideration. Thus, the Schr\"{o}dinger equation for the envelope function becomes\cite{datta}:
\begin{equation}
\begin{array}{r}
-\left( \begin{array}{cccc}
\hat{P}+\hat{Q}&0&-\hat{S}&\hat{R}\\
0&\hat{P}+\hat{Q}&\hat{R}^{*}&\hat{S}^{*}\\
-\hat{S}^{*}&\hat{R}&\hat{P}-\hat{Q}&0\\
\hat{R}^{*}&\hat{S}&0&\hat{P}-\hat{Q}
\end{array} \right)
{\bm \Psi}_{h}({\bm r})
\\
=E{\bm \Psi}_{h}({\bm r}),
\end{array}
\label{eq:holeSchrod}
\end{equation}
where
\begin{equation}
{\bm \Psi}_{h}({\bm r})=
\left( \begin{array}{c}\Psi_{lh\uparrow}({\bm r}),
\Psi_{lh\downarrow}({\bm r}),
\Psi_{hh\downarrow}({\bm r}),
\Psi_{hh\uparrow}({\bm r})
\end{array}
\right)^{T}.
\end{equation}
The indexes: $hh$ and $lh$ denote the components of envelope function for heavy holes and light holes, respectively. The symbols $\uparrow$ and $\downarrow$ distinguish bands related to opposite $z$-components of spin. The elements: $\hat{P},\hat{Q},\hat{R},\hat{S}$ of matrix in Eq. \ref{eq:holeSchrod} are given by the following formulae:
\begin{eqnarray}
\hat{P}&=&E_{V}({\bm r})\nonumber\\
& &+\alpha \left(
\frac{\partial}{\partial x}\gamma_{1}({\bm r})\frac{\partial}{\partial x}+
\frac{\partial}{\partial y}\gamma_{1}({\bm r})\frac{\partial}{\partial y}+
\frac{\partial}{\partial z}\gamma_{1}({\bm r})\frac{\partial}{\partial z}
\right),
\nonumber\\
\hat{Q}&=&\alpha \left(
\frac{\partial}{\partial x}\gamma_{2}({\bm r})\frac{\partial}{\partial x}+
\frac{\partial}{\partial y}\gamma_{2}({\bm r})\frac{\partial}{\partial y}-
2\frac{\partial}{\partial y}\gamma_{2}({\bm r})\frac{\partial}{\partial y}
\right),
\nonumber\\
\hat{R}&=&\alpha\sqrt{3}\left[
- \left(
\frac{\partial}{\partial x}\gamma_{2}({\bm r})\frac{\partial}{\partial x}-
\frac{\partial}{\partial y}\gamma_{2}({\bm r})\frac{\partial}{\partial y}
\right)\right.\nonumber\\
& & +\left. i \left(
\frac{\partial}{\partial x}\gamma_{3}({\bm r})\frac{\partial}{\partial y}+
\frac{\partial}{\partial y}\gamma_{3}({\bm r})\frac{\partial}{\partial x}
\right)
\right],
\nonumber\\
\hat{S}&=&\alpha\sqrt{3}\left[
\left(
\frac{\partial}{\partial x}\gamma_{3}({\bm r})\frac{\partial}{\partial z}+
\frac{\partial}{\partial z}\gamma_{3}({\bm r})\frac{\partial}{\partial x}
\right)\right.\nonumber\\
& & -\left. i \left(
\frac{\partial}{\partial y}\gamma_{3}({\bm r})\frac{\partial}{\partial z}+
\frac{\partial}{\partial z}\gamma_{3}({\bm r})\frac{\partial}{\partial y}
\right)
\right]
.\label{eq:PQRS}
\end{eqnarray}
Luttinger parameters $\gamma_{1}$, $\gamma_{2}$, $\gamma_{3}$ describe the non-isotropic effective mass of light and heavy holes, and $E_{V}$ is the position of the valence band top. 

Figure \ref{fig:spectrum} shows the spectrum of the 3D superlattice calculated by the plane-wave method. For the periodic material parameters: $m^{*}$, $E_{C}$, $\gamma_{1}$, $\gamma_{2}$,$\gamma_{3}$, $E_{V}$, Fourier transforms have been calculated analytically. The number of reciprocal lattice vectors in the expansion in a series of plane waves is set so as to obtain a satisfactory convergence of results. The dispersion relation is determined along the high-symmetry paths ${\rm \Gamma-K-M-\Gamma-A-H-L-A}$, ${\rm K-H}$ and ${\rm M-L}$ in the first Brillouin zone, shown in Fig. \ref{fig:geometry}c .

In the calculation of all the electronic and hole spectra we have assumed the following values of material parameters in the dots ($d=0$), the matrix ($d=0.35$) and the spacers ($d=1$)\cite{vurga}: $m^{*}= 0.067+0.083d$, $E_{C}=0.944d$, $\gamma_{1}=6.85-3.40d$, $\gamma_{2}=2.10-1.42d$, $\gamma_{3}= 2.90-1.61d$, $E_{V}= 1.519-0.75d$. The filling fraction of dots in the matrix in the quantum dot layer was fixed at $f=2\pi R^{2}/\sqrt{3}a^{2}=0.3$, and the lattice constant of the hexagonal superlattice at $a=50$\AA. The dot layer thickness $h_{2}$ and the spacer layer thickness $2h_{1}$ were adjusted so as to attain the limiting case of either isolated layers ($2h_{1}\gg h_{2}$) or infinite cylinders ($0\approx 2h_{1}\ll h_{2}$).

For quantum dots of typical size (see Fig. \ref{fig:spectrum}a) and layers of typical thickness only the lowest electron miniband is distinctly detached from the other minibands. The occurrence of this miniband as an intermediate extra level taking part in optical transitions between the valence band and the block of higher conduction minibands increases the efficiency of photon absorption. Besides the direct transition, available also in monolithic materials, between the valence band and the conduction band (VB-CB), a cascade transition valence band - intermediate band - conduction band (VB-IB-CB) becomes available as well. The power utilized by the photoconverter can be evaluated as:
\begin{eqnarray}
\begin{array}{l}
P_{out}=E_{G}\Big[I(E_{G},\infty)\Big.\\+\Big.\min\left(I(E_{G}-E_{I},E_{G}),I(E_{I},E_{G}-E_{I})\right)\Big],
\end{array}\label{eq:Pout}
\end{eqnarray}
where
\begin{eqnarray}
I(E_{1},E_{2})&=&2\pi(k_{B}T_{S})^{3}/h^{3}c^{2}\int_{\xi(E_{1})}^{\xi(E_{2})}\frac{\xi^{2}d\xi}{e^{\xi}-1},\nonumber\\
&&\xi(E)=E/k_{B}T_{s} \label{eq:flux}
\end{eqnarray}
denotes the flux of photons of energy ranging from $E_{1}$ to $E_{2}$; $E_{G}$ and $E_{I}$ are the width of the gap between VB and CB and the distance between the IB bottom and the CB top, respectively; $k_{B}$ is the Boltzmann constant, and $T_{S}$ is the temperature corresponding to the maximum of the solar spectrum. The ultimate efficiency of the solar cell is defined as the ratio of the utilized power $P_{out}$ to the power $P_{in}$ of the incoming photon flux:
\begin{eqnarray}
\eta=\frac{P_{out}}{P_{in}},\label{eq:effic}
\end{eqnarray}
where
\begin{equation}
P_{in}=2\pi^{5}(k_{B}T_{s})^{4}/15h^{3}c^{2},\label{eq:Pin}
\end{equation}
denotes the power of the flux of photons emitted by a black body at temperature $T_{S}$.

\section{Results}

\begin{figure}[!tbp]
\centering
\includegraphics[keepaspectratio, width = 3in]{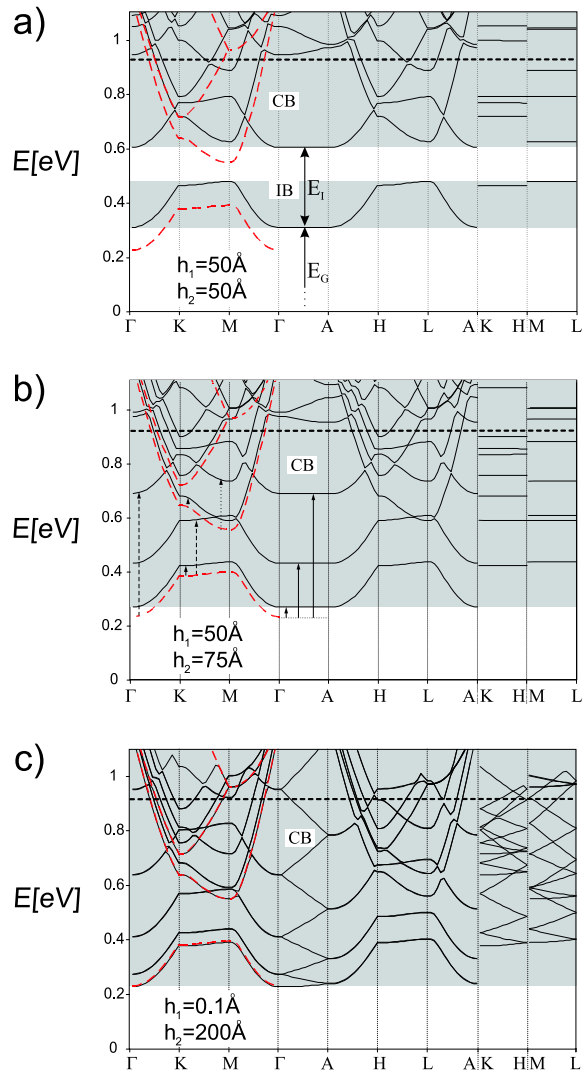}
\caption{The electronic spectra of the system presented in Fig. \ref{fig:geometry}, with GaAs rods (material A) and ${\rm Al_{0.35}Ga_{0.65}As}$ matrix (material B), for filling fraction $f=0.3$ and lattice constant $a=50$ {\AA} in the quantum dot layers. The spectra in (a) and (b) correspond to a system with almost isolated layers (spacer width $2h_{1}=100$\AA). The dispersive branches corresponding to the propagation along the rod axis are flat below the barriers in spacers (made of AlAs, material C). The higher number of modes resulting from a denser quantization along the rod axis (for longer rods, (b)) causes minigaps to close. The limit of infinite rods (c) is approached by using long rods and thin spacers ($2h_{1}=0.2$\AA). The parabolic dispersion relation is seen to fold in the first BZ for free motion along the rod axis. Red dashed line represents the dispersion relation in the 2D rod array model (when the motion along z-axis is not included.}\label{fig:spectrum}
\end{figure}

Figure \ref{fig:spectrum} shows the electronic spectra of the quantum dot superlattice depicted in Fig. \ref{fig:geometry}, calculated for different values of thickness $h_{2}$ of the dot layers and thickness $2h_{1}$ of the spacer layers. The spectra shown in Fig. \ref{fig:spectrum}a and b correspond to systems with relatively thick spacer layers ($2h_{1}=100$\AA), which implies the dot layers are isolated from one another. In this case electrons are bound within the layer and only allowed to propagate in the $x-y$ plane. Along the paths ${\rm\Gamma-A}$, ${\rm K-H}$ and ${\rm M-L}$ (propagation in the $z$ direction) the relation $E({\bm k})$ is nondispersive due to the restriction (binding) of the electron motion in the layer. Dispersion only occurs when the electron energy exceeds the barrier generated by the spacer (the barrier level is represented as a horizontal dashed line in Fig. \ref{fig:spectrum}). In a single isolated layer the electron motion in the layer plane (the $x-y$ plane) and that in the direction perpendicular to it (the $z$ direction) are nearly independent, as a consequence of the separability of the effective potential\cite{laz}: $E_{C}(x,y,z)=E_{C,x-y}(x,y)+E_{C,z}(z)$ (a strict separation of variables in (\ref{eq:duke}) requires the separation of variables also in the inverse effective mass\cite{klos}). Hence, the total electron energy is, approximately, the sum of the energy $E_{x-y}$ of the electron motion in the 2D periodic potential in the layer and the energy $E_{z}$ of the electron motion bound in a 1D well of width equal to the layer thickness. This is illustrated in Fig. \ref{fig:spectrum}, in which individual minibands are seen to form through a shift of minibands (represented as red dashed lines) corresponding to the motion in the 2D potential $E_{C,x-y}$ in the layer by the energy levels in the 1D potential well $E_{C,z}(z)$ (corresponding to the position of non-dispersive minibands on the path ${\rm\Gamma-A}$, where $k_{z}\neq 0$ and $k_{x-y}=0$). The effect is illustrated in Fig. 2b, in which the shift of the first and the second 2D minibands is indicated by dashed and dotted arrows, respectively; solid arrows indicate the energy levels $E_{z}$ corresponding to the non-dispersive minibands on the path ${\rm\Gamma-A}$.

The spectra presented in Fig. \ref{fig:spectrum}a and b correspond to structures of different quantum dot array thickness. As the height of dots increases, the number of bound states in the system grows as well. This translates into an increased number of minibands, and their denser coverage of the same energy range. As a result, for dot arrays thick enough (Fig. \ref{fig:spectrum}b) all the electronic minibands, including the lowest one, form a continuous block. For the considered structure to provide a basis of a high-efficiency solar cell, the dot array thickness should be limited so that the minigaps do not close.

Figure \ref{fig:spectrum}c shows the superlattice spectrum for strongly interacting dot arrays. As the spacers are very thin ($2h_{1}=0.2$\AA), the system can be regarded, with quite good approximation, as a superlattice with cylindrical rods of infinite length. In a 2D effective potential generated by this type of structure the electron motion along the z direction is free. This is evidenced by the parabolic dispersion relation along paths ${\rm\Gamma-A}$, ${\rm K-H}$ and ${\rm L-M}$, artificially folded to the first BZ. In an unfolded BZ the miniband structure for $k_{z}=0$ will be identical with that of the 2D superlattice (red dashed line).
\begin{figure}[!tbp]
\centering
\includegraphics[keepaspectratio, width = 2.5in]{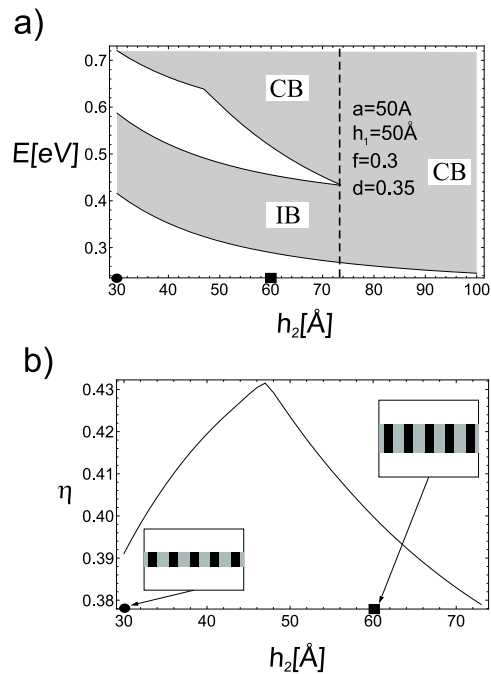}
\caption{ (a) The absolute electronic energy bands (the gray area) and gaps (white area) versus rod length $h_{2}$. (b) The ultimate efficiency versus $h_{2}$, for parameter values adjusted so as to apply the isolated layer approximation. The insets in (b) present the structures in cross section along rod axis.}\label{fig:efficiency}
\end{figure}

In order to estimate the ultimate efficiency of the solar cell we must know the distance $E_{I}$ between the bottom of the intermediate band (the first conduction miniband) and the bottom of the conduction band (the continuum formed by the minibands above the matrix potential), as well as the gap $E_{G}$ between the valence band and the conduction band (see Fig. \ref{fig:spectrum}a). We have determined both by investigating the position of absolute minigaps and minibands versus dot array thickness $h_{2}$ (see Fig. \ref{fig:efficiency}a). The calculations were performed for large spacer thickness, $2h_{1}=100$\AA, ensuring a sufficient isolation of the quantum dot arrays. The resulting plot of ultimate efficiency versus dot array thickness is shown in Fig. \ref{fig:efficiency}b.

All the minibands are easily seen to shift towards higher energies with decreasing $h_{2}$. This is caused by an increase in energy component $E_{z}$ related to the quantization in the $z$ direction. The efficiency of the solar cell is in this case mainly limited by the small shift $E_{I}$ of the intermediate band with respect to the conduction band. Thus, changes in width of the gap between the IB and the CB translate directly into changes in ultimate efficiency of the solar cell. The widest gap corresponds to dot array thickness $h_{2}\approx 50$ {\AA}  and implies maximum ultimate efficiency of the solar cell based on the structure under consideration. As evidenced in Fig. \ref{fig:spectrum}a, the maximum gap width entails the overlap of the two minibands originating from the first and the second minibands in the 2D dispersion relation. The gap closes, and the IB merges into the CB, for $h_{2}\approx 72$\AA, as a result of reduced spacing between energy levels $E_{z}$ and the large number of minibands in the system.

\section{Conclusion}
The thickness of the dot array has a substantial effect on the electronic spectrum of the structure considered in this study. With dot arrays thick enough, the electronic minibands caused by the in-plane periodicity are seen to close as a result of a large number of overlapping minibands. This is due to the dense quantization along the direction perpendicular to the layer. 
The efficiency of a solar cell using a quantum dot array, for structures based on AlGaAs, is mainly limited by the narrow minigaps in the conduction band. The ultimate efficiency decreases as the minigaps gradually close as a result of growing dot array thickness. The dot array thickness optimal from this point of view corresponds to the widest minigap; thus, for a hexagonal lattice of GaAs dots embedded in ${\rm Al_{0.35}Ga_{0.65}As}$ with filling fraction $f=0.3$, the optimal dot array thickness is $h_{2}=50$\AA, equal to the lattice constant of the 2D heterostructure.

\section*{Acknowledgements}
This study was supported by Polish Ministry of Science and Higher Education, grant No.~N~N507~3318~33.

\bibliographystyle{elsarticle-num}

\begin{thebibliography}{00}




\bibitem{3g2}M. A. Green, Third Generation Photovoltaics: Advanced Solar Energy Conversion, Springer, Berlin, 2006.
\bibitem{luque}A. Luque and A. Martí, Phys. Rev. Lett. {\bf 78}, 5014 (1997).
\bibitem{green}A. S. Brown, M. A. Green and R. P. Corkish, Physica E  {\bf 14} 212 (2002).
\bibitem{klos2}J. W. Kłos and  M. Krawczyk, J. Appl. Phys. {\bf 106}, 093703 (2009).
\bibitem{SQ}W. Shockley and H. J. Queisser, J. Appl. Phys.  {\bf 32}, 510 (1961)
\bibitem{tsaka}L. Tsakalakos, Mat. Sci. Eng. {\bf 62}, 175 (2008).
\bibitem{shao}Q. Shao, A. A. Balandin, A. I. Fedoseyev, and M. Turowski, Appl. Phys. Lett. {\bf 91}, 163503 (2007).
\bibitem{green2}S. Park , E. Cho, D. Song, G. Conibeer and M. A. Green, Sol. Energy Mater. Sol. Cells {\bf 93}, 684 (2009).
\bibitem{tomic}S. Tomic, T. S. Jones, N. M. Harrison, Appl. Phys. Lett.  
{\bf 93}, 263105 (2008).
\bibitem{fan}Z. Fan, H. Razavi, Jae-won Do, A. Moriwaki, O. Ergen, et al.,  Nature Mater. {\bf 8}, 648 (2009) .
\bibitem{krawczyk}M. Krawczyk, J. W. Kłos and A. Szał, Physica E {\bf 41}, 581 (2009). 
\bibitem{datta}S. Datta, Quantum Transport: Atom to Transistor, Cambridge University Press, 2005. 
\bibitem{duke}D. J. BenDaniel and C. B. Duke, Phys. Rev. 152 {\bf 683}, (1966).
\bibitem{vurga}I. Vurgaftman, J. R. Meyer and L. R. Ram-Mohan, J. Appl. Phys. {\bf 89}, 5815 (2001).
\bibitem{laz}O. L. Lazarenkova and A. A. Balandin, J. Appl. Phys.  {\bf 89}, 5509 (2001). 
\bibitem{klos}J. W. Kłos and M. Krawczyk,  Materials Science-Poland {\bf 26}, 970 (2008) .

 \end{thebibliography}



\end{document}